\documentclass[sigconf,anonymous=false]{acmart}

\def\BibTeX{{\rm B\kern-.05em{\sc i\kern-.025em b}\kern-.08emT\kern-.1667em\lower.7ex\hbox{E}\kern-.125emX}}
    
\usepackage[binary-units=true]{siunitx}
\usepackage{amsfonts}

\usepackage{amsmath,amssymb}
\usepackage{ulem}
\usepackage{units}
\usepackage{multicol}
\usepackage{fancyhdr}
\usepackage{algorithm}
\usepackage{algpseudocode}
\usepackage{enumitem,kantlipsum}
\copyrightyear{2019} 
\acmYear{2019} 
\setcopyright{acmcopyright}
\acmConference[CF '19]{Proceedings of the 16th conference on Computing Frontiers}{April 30-May 2, 2019}{Alghero, Italy}
\acmBooktitle{Proceedings of the 16th conference on Computing Frontiers (CF '19), April 30-May 2, 2019, Alghero, Italy}
\acmPrice{15.00}
\acmDOI{10.1145/3310273.3322822}
\acmISBN{978-1-4503-6685-4/19/05}
\newcommand{\secref}[1]{Sec.~\ref{#1}}
\newcommand{\figref}[1]{Fig.~\ref{#1}}
\newcommand{\tabref}[1]{Table~\ref{#1}}
\newcommand{\Algref}[1]{Alg.~\ref{#1}}
\newcommand{\Alglineref}[1]{line~\ref{#1}}
\newcommand{\Eqref}[1]{Eq.~\ref{#1}}

\newcommand{\norm}[1]{\left\lVert#1\right\rVert}
\newcommand{\Exp}{\mathbb{E}}

\begin{document}
\title{Embedding Principal Component Analysis for Data Reduction in Structural Health Monitoring on Low-Cost IoT Gateways}

\author{Alessio Burrello}
\authornote{Both authors contributed equally to this research.}
\email{alessio.burrello@unibo.it}
\author{Alex Marchioni}
\authornotemark[1]
\email{alex.marchioni@unibo.it}
\affiliation{%
  \institution{DEI, University  of  Bologna}
  \city{Bologna}
  \state{Italy}
  \postcode{40136}
}

\author{Davide Brunelli}
\email{davide.brunelli@unitn.it}
\affiliation{%
  \institution{DII, University  of  Trento}
  \city{Trento}
  \state{Italy}
  \postcode{38123}
}
\author{Luca Benini}
\email{lbenini@iis.ee.ethz.ch}
\affiliation{%
  \institution{DEI, University  of  Bologna}
  \city{Bologna}
  \state{Italy}
  \postcode{40136}
}
\affiliation{%
  \institution{IIS, ETH Zurich}
  \city{Zurich}
  \state{Switzerland}
  \postcode{8092}
}

\begin{abstract}

Principal component analysis (PCA) is a powerful data reduction method for Structural Health Monitoring. 
However, its computational cost and data memory footprint pose a significant challenge when PCA has to run on limited capability embedded platforms in low-cost IoT gateways.
This paper presents a memory-efficient parallel implementation of the streaming History PCA algorithm. 
On our dataset, it achieves 10$\times$ compression factor and 59$\times$ memory reduction with less than 0.15 dB degradation in the reconstructed signal-to-noise ratio (RSNR) compared to standard PCA. 
Moreover, the algorithm benefits from parallelization on multiple cores, achieving a maximum speedup of 4.8$\times$ on Samsung ARTIK 710.
\end{abstract}
\keywords{Embedded platforms, Streaming PCA, Structural Health Monitoring, Edge computing, IoT}

\maketitle
\section{Introduction}
\thispagestyle{fancy}
\fancyhf{}
\chead{Published as a conference paper at ACM International Conference on Computing Frontiers 2019}
The Internet of Things (IoT) envisions billions of devices that can sense, compute and potentially communicate with users or among them (machine-to-machine)~\cite{IoT}.
The IoT involves large sensors and devices networks and poses new challenges in finding innovative and scalable approaches to collect and to process the potentially huge amount of data.
A very promising IoT application is Structural Health Monitoring (SHM)~\cite{SHMIot1,SHMIot2} whose aim is to provide information about the building conditions.
To improve the maintainability of large structures such as space vehicles~\cite{Sens_net2}, masonry buildings~\cite{Sens_net1}, or bridges~\cite{Sens_net_bridges}, large sensors networks are installed and supervised by local gateways.
These monitoring systems continuously acquire and transmit data to a central unit for storage and processing purposes.
Depending on the scenario, a single gateway manages the data flow from a group of sensors, with a possible focus on either the sensors-gateway communication, as in \cite{Pareschi17, Bortolotti18}, or on the link from the gateway to the main processor \cite{Mangia18_IoTJ, Mangia18_JETCAS}.

Even though the role of the central unit is often played by a cloud platform~\cite{CloudIoT}, the distribution of the processing to the edge of the cloud, i.e. by including gateways and sensors in the data processing chain, has demonstrated several advantages such as improved security, reduced latency and lower costs~\cite{Shi16}.

In this framework, reducing the cloud storage space and the network traffic from the gateway to the cloud is extremely useful in many application scenarios \cite{Aazam14}. 
Therefore, a compression algorithm running on the gateway is required.
%
Among compression approaches, methods based on the Principal Component Analysis~\cite{Jolliffe1986} (PCA) are drawing attention in the processing of large dataset~\cite{PCA_Proceedings}. 
For these class of methods, the dimensionality reduction is achieved by exploiting the correlation between data features.
As a counterpart, an accurate estimation of the signal correlation requires greater memory and computational resources compared to other techniques~\cite{Mangia12}. 
This gap is partially filled by the recently emerging streaming approaches~\cite{HPCA, Li16, Hardt2014, Mitliagkas13}, that, with a lower memory footprint, exhibit a high accuracy in the correlation estimation.
Particularly, the History PCA~\cite{HPCA} is demonstrated to outperform other streaming approaches. 
In details, the History PCA has already been tested on four different large-scale public datasets (NIPS and NYTimes from UCI data, and RCV1, KDDB from LIBSVM data sets as presented in~\cite{HPCA}) achieving a lower approximation error compared to other algorithms.
History PCA achieves an approximation error (difference between the computed eigenvalues and the PCA eigenvalues) lower than $10^{-6}$, which is well suited for many tasks like anomaly detection or spectral analysis.

Tests conducted in~\cite{PCA_spatial_traffic1,PCA_spatial_traffic2} explore PCA solutions to reduce the traffic load exploiting spatial correlation among sensors.
One alternative solution to enhance the compression is to exploit the autocorrelation of the single sensor time series. 
This allows to treat the sensor as a unique entity, running a single PCA instance for each sensor.

In this paper we tackle the challenge of embedding PCA compression on low-cost IoT gateways, and we address two key limitations of these HW platforms, namely: limited on-board memory and low computational power. 
We describe the following contributions:
\begin{itemize}[leftmargin=*,topsep=0pt]
    \item We present the application of the History PCA on a real-life SHM problem, i.e. the monitoring of a viaduct. To the best of our knowledge, this is the first attempt to embed the training of a Principal Component Analysis algorithm on a large scale sensor network on a low-cost gateway. We efficiently train our algorithm on a single-sensor temporal series and we demonstrate the suitability of the gateway to manage a 45 sensor (3-axis accelerometers) network.
    \item Our HPCA implementation on an ARTIK 710 Module~\cite{Artik710} with 8 cores achieves 4.8$\times$ training speed-up and 2.1$\times$ energy saving compared to its single core execution. We also demonstrate that the ARTIK 710 is superior to Raspberry Pi 3~\cite{RPI3}, reaching 2.6$\times$ higher speed-up with 1.2$\times$ energy reduction. 
    \item We further investigate how different configurations of the History PCA could fit different memory constraints, showing that we can save up to $59\times$ memory.
\end{itemize}

The rest of the article is organized as follows: \secref{sec:algorithm} presents the PCA and the History PCA algorithm.
\secref{sec:dataset_platforms} introduces our dataset and describes the platforms.
Finally, \secref{sec:results} discusses the experimental results, while \secref{sec:conclusion} concludes the paper with final remarks.

\section{History PCA}
\label{sec:algorithm}

Given a dataset $X \in \mathbb{R}^{d \times N}$ with $d$ features and $N$ instances, the PCA consists in the eigen decomposition of the correlation matrix 
\begin{equation}
C_x = \frac{1}{N} X X^\intercal = Q \Lambda Q^\intercal
\label{eq:corr}
\end{equation}
where $Q\in \mathbb{R}^{d \times d}$ is an orthonormal matrix containing the column eigenvectors of $C_x$ and $\Lambda = \rm{diag}(\lambda_1, \dots, \lambda_d)$ is the eigenvalues matrix with $\lambda_1 \ge \lambda_2, \ge \dots, \ge \lambda_d$.

Dimensionality reduction is obtained by projecting the signal onto the $k$ eigenvectors corresponding to the higher eignevalues. The average reconstruction error is the sum of the eigenvalues corresponding to the eigenvectors not involved in the projection. Let $x \in \mathbb{R}^d$ be a signal instance, and $\hat{x}$ its reconstruction, the compression error is computed as:
\begin{equation}
\Exp_x\left[\norm{x - \hat{x}}_2\right] = \sum_{j=k+1}^d \lambda_j
\label{eq:recerr}
\end{equation}

In this paper we apply the PCA-based compression to time series, by representing signals as non-overlapping windows of $d$ subsequent samples. 
Firstly, $N$ signal instances are collected in order to estimate the correlation matrix $C_x$ (\Eqref{eq:corr}). As a result, by following \Eqref{eq:recerr}, it is possible to fix the average compression error and determine the number of top eigenvectors involved in compression, $k$. 
Then, each next incoming signal instance $x \in \mathbb{R}^d$ is compressed by $y = {Q^{(k)}}^\intercal x$, where $Q^{(k)} \in \mathbb{R}^{d \times k}$ is the matrix with the top $k$ egenvectors. Eventually, signal recovery is obtained by $\hat{x} = {Q^{(k)}} y$.

The classical approach entails to simultaneously store all the $N$ windows and then perform the PCA analysis, thereby demanding for a large memory array. In contrast, the edge computing platforms are tightly constrained by limited memory space.
It is therefore interesting to adopt streaming approaches~\cite{HPCA,Li16, Oja1982,Hardt2014,Mitliagkas13}, which follow an incremental scheme to update the eigenvectors estimate every new block of $B$ instances, with $B \ll N$. 
Among them, History PCA (HPCA)~\cite{HPCA} has recently emerged. Based on the block stochastic power method \cite{Mitliagkas13}, HPCA aims to improve accuracy by using more information about past signal instances.

\begin{algorithm}
\caption{\label{alg:HPCA} HPCA}
\begin{algorithmic}[1]
\State Input: ${X_1, \dots, X_n}$, block-size: $B$.
\State $S_0^{(i)} \sim N(0,I_{d \times d}), 1 \le i \le k$
\State $Q_1 \gets QR_{decomposition}(S_{0}) $
\For{$i \gets 1, \dots, m$} 
    \State $S_1 \gets Q_1 + \frac{1}{B} X_1 X_1^\intercal Q_1$
    \State $Q_1, \cdot \gets QR_{decomposition}(S_1)$
\EndFor
\State $\lambda_j \gets \norm{S_1[:,j]}_2$ for $j = 1, \dots, k$
\State $\Lambda_1 \gets diag(\lambda_1, \dots, \lambda_k)$
\For{$\tau \gets 2, \dots, n$}
    \State $Q_{\tau} \gets Q_{\tau-1}$
    \For{$i \gets 1, \dots, m$}  
        \State $S_{\tau} \gets \frac{\tau-1}{\tau} Q_{\tau-1} \Lambda_{\tau-1} Q_{\tau-1}^\intercal Q_{\tau} + \frac{1}{\tau} \frac{1}{B} X_{\tau} X_{\tau}^\intercal Q_{\tau}$ \label{alg:HPCA:core}
        \State $Q_{\tau}, \cdot \gets QR_{decomposition}(S_{\tau})$ \label{alg:HPCA:qr}
    \EndFor
    \State $\lambda_j \gets \norm{S_{\tau}[:,j]}_2$ for $j = 1, \dots, k$
    \State $\Lambda_{\tau} \gets diag(\lambda_1, \dots, \lambda_k)$
\EndFor
\State Output: $Q_n$
\end{algorithmic}
\end{algorithm}

The HPCA method is provided in \Algref{alg:HPCA} and it is deeply explained in \cite{HPCA}.
The algorithm runs a new step $\tau$ every time a block $X_\tau \in \mathbb{R}^{d \times B}$ is gathered updating the top $k$ eigenvectors estimate as columns of $Q_\tau \in \mathbb{R}^{d \times k}$. After $n=N/B$ steps the method returns the final estimate $Q_n$.

The core of the algorithm is contained in \Alglineref{alg:HPCA:core}, which represents a generalization of the classical power method ($\omega_{\tau} \gets \frac{1}{B} X_{\tau} X_{\tau}^\intercal \omega_{\tau-1}$ with $\omega$ as top eigenvector estimate). The second term of the addition represents the power method of the current block whereas the first contains the information extracted from the previous blocks; the two $\tau$-based coefficients weight the terms so that each block equally contributes to the final estimate. Besides, the QR decomposition of \Alglineref{alg:HPCA:qr} is necessary to ensure the orthonormality among the eigenvectors estimate. The repetition of these steps for $m$ times ensures the estimate to quickly converge to the actual eigenvectors with good accuracy \cite{HPCA}.

The computational cost of the HPCA is dominated by the matrix multiplications $O(dk(k+B))$ and by the QR decomposition $O(dk^2)$.
The memory occupancy is $\sim O(d(k+B))$, implying a gain of $\nicefrac{N}{(B+k)}$ with respect to the classical PCA. With $B=1$ the algorithm reaches its lowest memory footprint. 

These values are consistent with other streaming PCA algorithm known in literature \cite{Oja83, Li16, Mitliagkas13} which present complexity $O(dk^2)$ and $O(dk)$ memory. The advantages of HPCA reside in the robustness in the parameter tuning as shown in Section~\ref{sec:use_case_results} and the improved rate of convergence granted by the iteration of the internal loop as demonstrated in \cite{HPCA}.

Summarizing, the HPCA algorithm is characterized by the following parameters: 
\begin{itemize}
    \item signal dimension $d$: a high value of $d$ allows greater compression ratio, but introduces a delay equal to $\nicefrac{d}{f_s}$ that could not meet real-time requirements.
    \item rank $k$: the number of eigenvectors to estimate. It influences the compression ratio $\rm{CR} = \nicefrac{d}{k}$ and, consequently, the compression quality.
    \item block size $B$: number of signal instances used in each step of HPCA algorithm.
    \item number of repetitions of the HPCA internal loop $m$.
\end{itemize}

The computational cost of the HPCA is dominated by the matrix multiplications $O(dk(k+B))$ and by the QR decomposition $O(dk^2)$.
Besides, the memory occupancy is $\sim O(d(k+B))$, implying a gain of $\nicefrac{N}{(B+k)}$ with respect to the classical PCA. With $B=1$ the algorithm reaches its lowest memory footprint.

\section{Dataset \& Platforms overview}
\label{sec:dataset_platforms}
In this section, we describe the dataset, and the structural health monitoring installation. Then, we introduce the Artik 710 module, and the Raspberry Pi 3 platforms. 
We have chosen these two platforms as representatives of the last generation of gateways for edge computing.
\subsection{Dataset}
Our experiments target a structural health monitoring (SHM) installation on a viaduct which is monitored by 90 sensor nodes equipped with a 3-axis accelerometer, temperature and humidity sensor. 
The nodes sense the vibrations of the internal tendons in different sections with sample frequency $f_s = \unit[100]{Hz}$. The readings are gathered by two gateways which pack the data as 32-bit integer and send it to the cloud for storage and analysis purposes.
The PCA-based compression is used to reduce bandwidth and storage space on the cloud and it is applied independently to each time series produced by each acceleration axis.

For our analysis, we use a dataset comprising a single-axis time series of the accelerometer mounted on a sensor node. We consider two traces, each one \unit[12]{hours} long, acquired in two different days. The former is used to estimate the top $k$ eigenvectors while the latter is needed to measure the performances in terms of quality of reconstruction. As a figure of merit, we use the Reconstruction Signal to Noise Ratio (RSNR) defined as:
$$ \rm{RSNR} = \frac{\norm{x}_2^2}{\norm{x-\hat{x}}_2^2} $$
where $x$ is the signal instance and $\hat{x}$ is the reconstructed instance after compression.
\subsection{Platforms overview}
\label{sec:platform}
\subsubsection{Raspberry Pi 3}
The Raspberry Pi 3 Module B \cite{RPI3} (Rpi3) is a single-board computer initially developed for teaching application. 
Now, it is actively used in many fields such as robotics, smart sensor control, and structural health monitoring.
The board comprises the Broadcom BCM2837 SoC, equipped with a 1.2 GHz 64-bit 4-core Cortex A-53, and 1GB low power DDR2 clocked at 900 MHz. 
\subsubsection{Samsung Artik 710}
The Samsung ARTIK 710 Module~\cite{Artik710} is an embedded computing System-in-Module targeted at high-end gateways with local processing and analytics.
It consists of a 8-core 64-bit ARM Cortex-A53 running at 1.4 GHz with 256KB shared L2-Cache, and two 512MB DDR3 16-bit memory chips with 32-bit memory interface, which provides a throughput of 6.4 GB/s.

We use the default power mode for all our experiments on the Rpi 3 and on the Artik 710 Module, and an external Keithley 2400 SourceMeter SMU for power measurements.

\section{Results}
\label{sec:results}
The History PCA is implemented using optimized Numpy Python 3.5 library, relying on highly optimized BLAS and LAPACK libraries for linear algebra computation.
In the following, we first present the reconstruction error and the tuning of parameters of the HPCA algorithm on our dataset (\secref{sec:use_case_results}).
Then we analyze memory occupation, execution time (along with the parallelization speed up), and energy consumption of the algorithm, on both the ARTIK 710 and the Rpi 3 introduced in \secref{sec:platform} (\secref{sec:hw_results}).

\subsection{Use case: Vibration based SHM}
\label{sec:use_case_results}
To evaluate the performance of HPCA on our dataset, we fix $d$ to \unit[500]{samples} (corresponding to \unit[5]{s}) and the compression quality to \unit[20]{dB}. We consequently set $k=50$ ($\rm{CR} = 10$).
The parameters $B$ and $m$ are tuned to achieve the best performance. 

\begin{figure}
  \centering
\includegraphics[width=1\columnwidth]{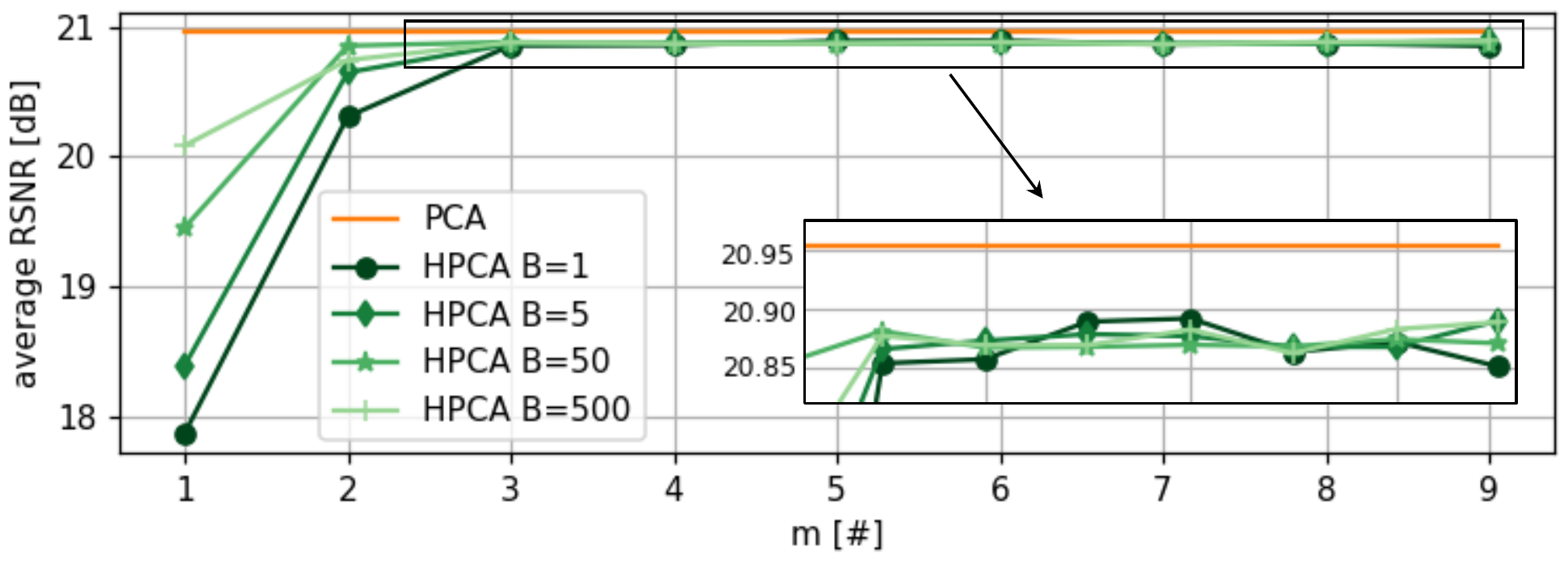}
  \caption{Average RSNR depending on the number of HPCA internal loops $m$ for different values of block-size $B$ with $N=8650$ (\unit[12]{hours}), $d=500$, $k=50$.}
\label{fig:tuning}
\end{figure}
\begin{figure}
  \centering
\includegraphics[width=1\columnwidth]{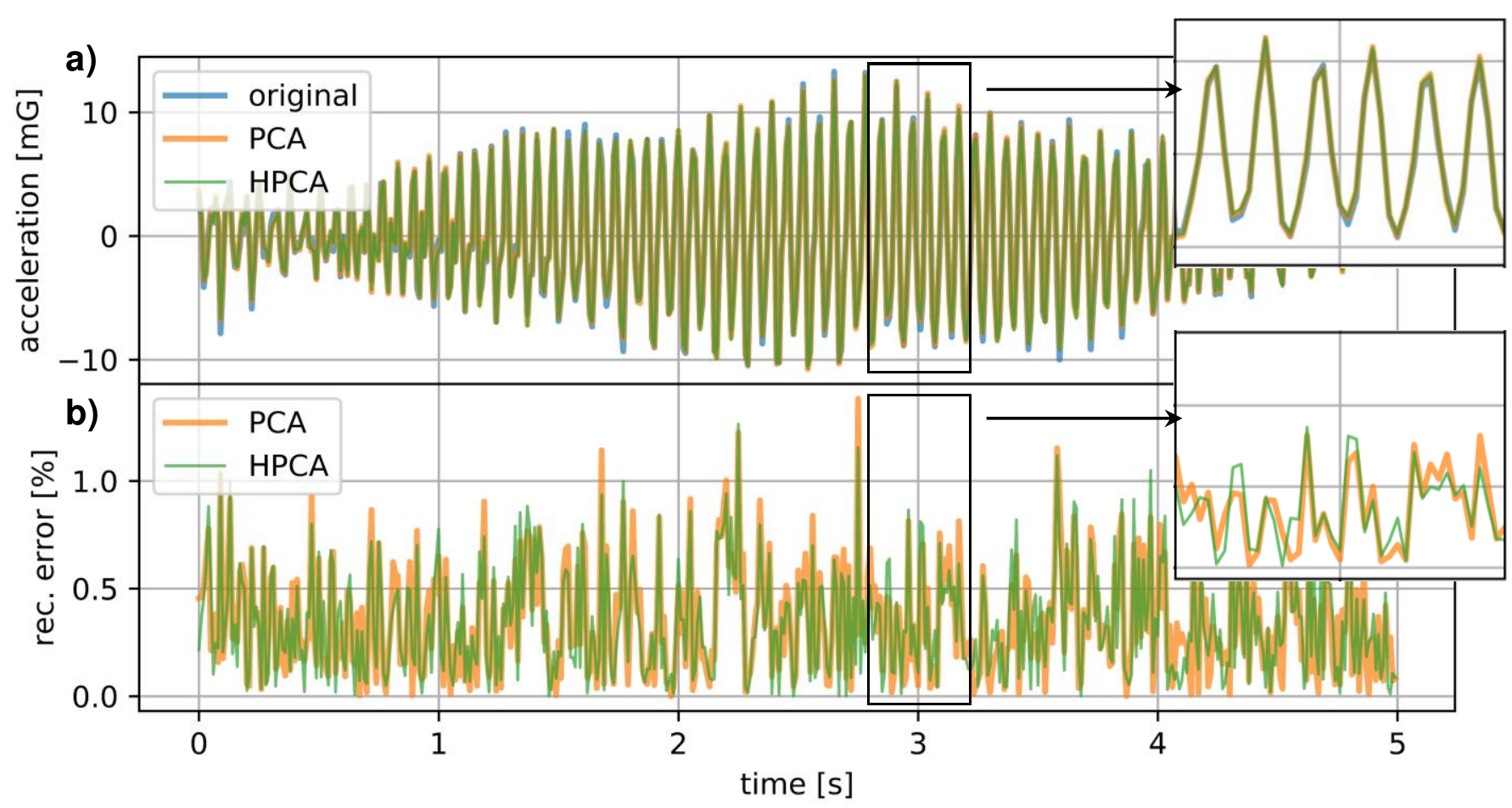}
  \caption{In a) a signal instance compared with its reconstructions after compression based on both PCA (RSNR=20.39dB) and HPCA (RSNR=20.31dB). b) shows the reconstruction errors relative to the RMS value of the signal instance.}
  \label{fig:cisa_rec}
\end{figure}
\figref{fig:tuning} depicts the average RSNR with varying $B$ and $m$. For low values of $m$ the HPCA suffers from up to \unit[3]{dB} RSNR loss compared to PCA. Lower values of $B$ emphasize the performance gap. 
By increasing $m$ to 3 or higher values, the RSNR difference reduces to just \unit[0.15]{dB} and becomes independent of the block-size $B$. This result highlights the HPCA robustness to parameters variation and it represents the main advantage of the HPCA with respect to the other methods \cite{Oja83, Li16, Hardt2014, Mitliagkas13} which need a fine parameter tuning.

We therefore fix $m = 3$ and $B = 50$.

On our test dataset, the HPCA achieves an average RSNR = \unit[20.88]{dB}, almost matching the PCA performance (average RSNR = \unit[20.95]{dB}).
As we expect, varying the value of $B$ from 1 to 500 does not impair the HPCA performance (RSNR $\in[\unit[20.85]{dB},\unit[20.90]{dB}]$).
In \figref{fig:cisa_rec} we report an example of signal instance and its reconstructions after compression. Noteworthy, the punctual reconstruction error (defined as $\nicefrac{(x-\hat{x})}{\norm{x}_2}$) of HPCA and PCA are highly correlated, meaning that the vectors identified by the two algorithms are practically the same.

\subsection{Rpi \& Artik Measurements}
\label{sec:hw_results}
\begin{table}[t]
\centering
\small
\caption{Performance of History PCA on Artik 710 versus Raspberry Pi 3. Results refer to the execution with
d = 5000, B = 1, k = 500, m=3; MM, QR stand for matrix multiplication and QR decomposition.}
\label{table:Energy}
\begin{tabular}{rrrrr}
\toprule
\textbf{CORES [\#]} & 1  & 2 & 4 & 8   \\\midrule
\multicolumn{5}{@{}l}{\textbf{Samsung ARTIK 710}}\\
time [\SI{}{\second}] & 50.7 (1$\times$) & 28.0 (1.8$\times$) & 15.6 (3.3$\times$) & 10.5 (4.8$\times$)   \\
MM [\SI{}{\second}] & 29.8 (1$\times$) & 16.5 (1.8$\times$) & 9.3 (3.2$\times$)  & 5.4 (5.5$\times$)   \\
QR [\SI{}{\second}] & 19.8 (1$\times$) & 11.4 (1.7$\times$) & 6.7 (3.0$\times$)  & 5.1 (3.9$\times$)   \\
\midrule
\multicolumn{5}{@{}l}{\textbf{Rasberry Pi 3 Model B}}\\
time [\SI{}{\second}]     & 59.7 (1$\times$) & 34.8 (1.7$\times$) & 23.1 (2.6$\times$)  & n.a.  \\
MM [\SI{}{\second}]     & 37.7 (1$\times$) & 19.9 (1.9$\times$) & 12.6 (3.0$\times$)  & n.a.   \\
QR [\SI{}{\second}]    & 21.6 (1$\times$) & 14.6 (1.5$\times$) & 12.7 (1.7$\times$)  & n.a.   \\
\bottomrule
\end{tabular}
\end{table}
To measure the execution time, the parallelization speed-up and the energy consumption of the algorithm on the two platforms, we consider a single step of the algorithm, i.e. an iteration of the $\tau$-loop.
To discuss the trade-off between memory and energy we execute a full pass through all the training data. 

In \tabref{table:Energy}, we analyze the timing and the speed up of the algorithm on both the ARTIK 710 and the Rpi3.
For this experiment, we decide to increase $d$ to 5000 and maintain a CR of 10$\times$, extending our results to a more general scenario with a very high number of features.
With this setting, one HPCA step requires \unit[50.7]{s} (\unit[59.7]{s}) to run on single-core ARTIK 710 (Rpi3) and benefits of up to 4.8$\times$ (2.6$\times$) speed-up from parallelization.
Note that the speed-up is strongly limited by the QR decomposition, that nearly accounts for 50\% of the execution time in the max-core configuration of both platforms.
Indeed, the QR decomposition relies on many sequential steps, which implies a lower speed-up gain from multicore execution (3.9$\times$/1.7$\times$ on Artik710/Rpi3).  
Furthermore, the lower speed up achieved by the Rpi3 with equal parallelization (2.6$\times$ vs 3.3$\times$ with 4-cores) shows that the application is memory bound by the DDR2 lower throughput. 

\begin{figure}
  \centering
\includegraphics[width=1\columnwidth]{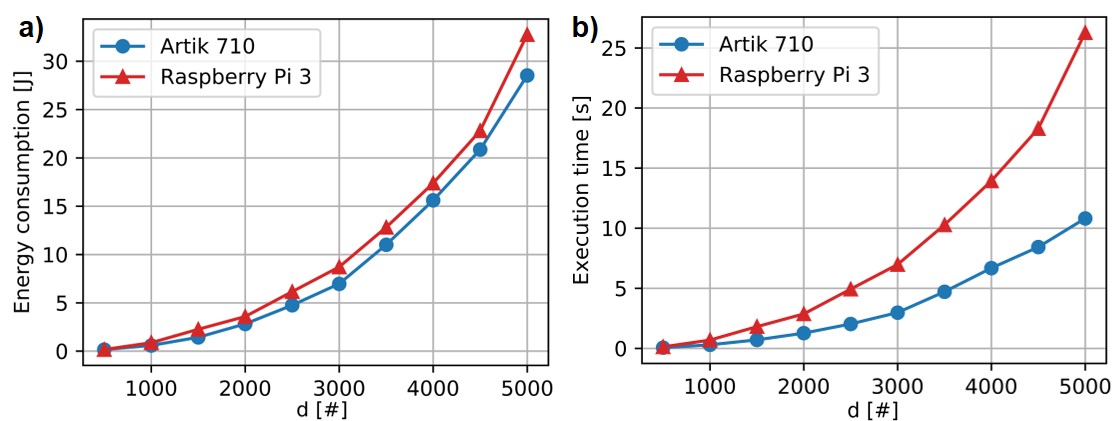}
  \caption{ a) energy and b) time comparison of  HPCA running on Artik 710 versus Raspberry Pi 3 (maximum core configuration) with varying number of features ($d$), $k = \nicefrac{d}{10}$, $B=1$ and $m=3$.   }
  \label{fig:feat_ener_time}
\end{figure}

\figref{fig:feat_ener_time} further analyzes these aspects for different features' numbers. 
As expected, both energy and time dramatically increase with the number of features (the complexity is $\sim dk^2$, with $k=\nicefrac{d}{10}$).
Moreover, the gap between execution time on Artik 710 and Rpi3 increases with the number of features, confirming that the DDR2 accesses limit the speed of the HPCA execution on Rpi3. 
With $d=5000$ the execution of the HPCA is 2.6$\times$ faster on the ARTIK 710 and achieves 1.2$\times$ energy saving.
%
Overall, the parallelized version of the HPCA on the Artik platform is executed in \unit[21]{\%} of the total time needed for training with $d=5000$ (\unit[10.5]{s} every \unit[50]{s} of signal acquisition), fitting well the real-time constraint given by the 100 Hz sampling frequency; decreasing $d$ to 500 reduces the percentage to only \unit[0.8]{\%}, allowing the simultaneous training of more sensors on a single gateway.

We also evaluate the scalability of the algorithm by increasing the size of the block B with fixed value of $k$ and $d$, respectively, 50 and 500. 
Note that a lower value of $B$ implies lower memory requirement (memory $\sim \mathcal{O}(d(k+B))$), at the expense of a higher number of steps $n$ ($n = N/B$).
More in details, the memory complexity of the HPCA algorithm is represented by 
$$ memHPCA = data_{size}\times(3\times d\times k + d\times B + k\times k)$$
where $d\times k$ and $k\times k$ represent the occupation of the intermediate data to compute the $Q_{\tau}$ matrix and $d\times B$ represents the input $X_i$ block. Hence, for $B \gg k$, the memory occupation depends linearly on $B$. On the other hand, increasing the value of $B$ reduces the number of $X_i$ blocks, and consequently the number of steps required by the HPCA to process the whole data needed for the training.
For instance, by doubling the block size, HPCA requires 10-20\% more CPU-time per step, but the number of execution step drops by 50\%, i.e., an increase of $B$ correspond to a decrease in the energy required to complete the training.

\begin{figure}
  \centering
\includegraphics[width=1\columnwidth]{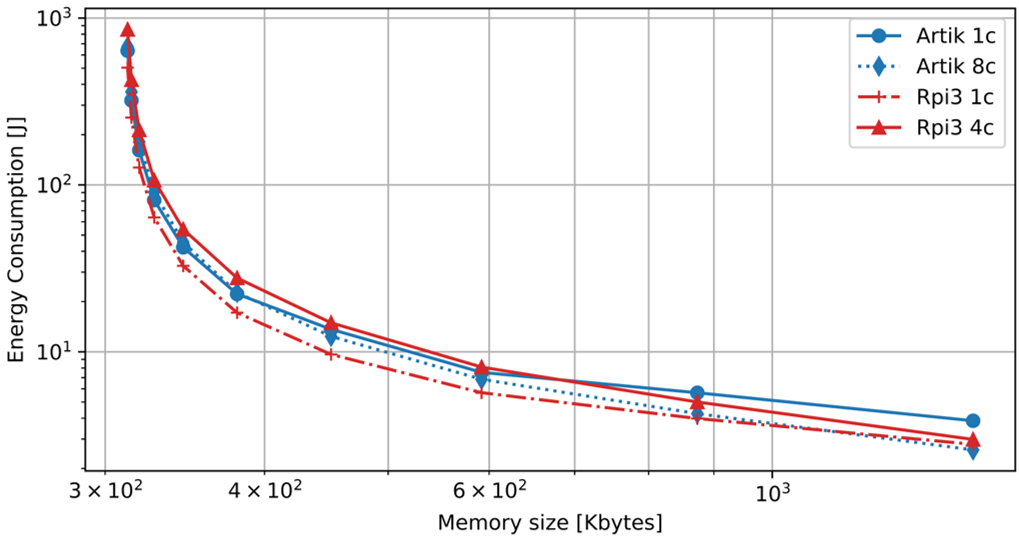}
  \caption{Trade-off between energy consumption and memory occupation on Artik 710 and Raspberry Pi 3 after a full pass over all the training data of the HPCA. Settings: $d=500$, $k=50$, $B\in[1,512]$, $m=3$.  }
  \label{fig:mem_energy}
\end{figure}

\figref{fig:mem_energy} shows the trade-off between necessary memory and energy consumption. 
As previously exposed, for high values of B ($B \gg k$) the storage for the data block $X_\tau$ ($X_\tau \in \mathbb{R}^{d \times B}$) dominates the memory occupation, while for $B \ll k$, it is determined by the matrices $S$ and $Q_\tau$ ($S,Q_\tau \in \mathbb{R}^{d \times k}$).
By increasing $B$ from 1 to 50, the HPCA necessitates $1.5\times$ memory (from 312 KB to 420 KB) and saves almost 50$\times$ energy consumption on both the platforms, whereas still increasing B to 512 causes further $3\times$ memory occupation with only 5$\times$ more energy saving.
%
Overall, setting $B=1$ (minimum memory footprint) allows a 59$\times$ memory reduction with respect to standard PCA, moving from 18.3 MB to 312 KB.
Moreover, managing our 45 sensor installation with 3 axes time series requires only 42.1 MB, with respect to the 2.5 GB needed by PCA.

\figref{fig:mem_energy} also depicts the different energy consumption between the two platforms. 
On a single-core, HPCA is compute bound and the Artik module shows higher energy consumption due to the Artik DDR3 RAM which is more power-hungry than the DDR2 mounted on Rpi3. 
When we move to the multi-core execution, the bottleneck becomes the memory access. The DDR3 in the Artik module grants a quicker access to the RAM that allows the 8 cores to run more efficiently, resulting in lower power consumption with respect to the single-core configuration. Conversely, the DDR2 mounted on RPi3 does not allow the 4 cores to exploit the maximum speed-up since most of the time is spent in load-store operations and the speed-up is not able to compensate the higher consumption due to the multi-core execution.


\section{Conclusion}
\label{sec:conclusion}
This work presents accelerating History PCA on Artik 710 and on Raspberry Pi 3, for data compression in a structural health monitoring system.
We obtained the same performance of the standard PCA ($10\times$ compression factor with \unit[20]{dB} of average RSNR), with 59$\times$ lower memory footprint.
We also compare the execution of the algorithm on the two platforms for different values of $d$, demonstrating that Artik 710 achieves 2.6$\times$ faster execution and 1.2$\times$ energy reduction, with respect to Raspberry Pi 3.

Our future work will focus on moving the computation of the HPCA on the single sensor under even tighter memory and computational effort constraints.

\section{Acknowledgments}
The research contribution presented in this paper has been funded by a research grant of ST Microelectronics and by the Emilia Romagna region Doctoral Program.
\bibliographystyle{ACM-Reference-Format}


%

\end{document}